\def\gsim{\agt}
\def\lsim{\alt}
\def\noi{\noindent}

\def\beq{\begin{equation}}
\def\eeq{\end{equation}}

\def\gg{\gamma\gamma}
\def\g{$\gamma$}
\def\ll{$ll$}

\documentstyle[preprint,revtex]{aps}
\begin{document}
\draft
\preprint{TRI-PP-92-124}
\preprint{Dec. 16, 1992}
\begin{title}
What can the L3 $\gamma\gamma ll$ events be?
\end{title}
\author{Robert Garisto and John N. Ng}
\begin{instit}
\begin{center}
TRIUMF,
4004 Wesbrook Mall,
Vancouver, B.C.,
V6T 2A3,
Canada
\end{center}
\end{instit}
\begin{abstract}
We consider the 4 \g\g\ll\ ($l=\mu,\ e$) events reported by the
L3 collaboration, and go through the logical possibilities which
could explain the events.
If they are not coincidental bremsstrahlung events, we find that
the physics which they could point to is extremely limited.  One
possibility would be to have a new 60 GeV scalar (or pseudoscalar)
particle $X^0$ with an off-diagonal coupling to a $Z$ and $Z'$ which is
non-perturbative ($\alpha > 1$), where the $Z'$ couplings to $\nu\bar\nu$
are suppressed. One could also construct a model involving $X^0$, and
a second scalar $X'^0$ with a large $X'll$ coupling.
We do not promote either of these models, but
hope they would prove to be useful guidelines, should the L3 events
turn out to be new physics.
\end{abstract}



\section {Introduction}

Recently, the L3 collaboration at LEP reported four unusual \g\g\ll\ events
(there are three events with $l= \mu$ and one with $l=e$)
with an $M_{\gg} \simeq 60$GeV,
in a sample of $10^6$ $Z$ decays \cite{Lthree}.
Though it is unusual to have bremsstrahlung events with such a high
invariant mass, the lepton-photon angles are quite small
for three of the
four events, and the L3 collaboration estimates that the probability
of having such events coming from bremsstrahlung
due to chance is as large as 0.1\%
\cite{Lthree}.

For the purposes of this paper, we will assume that
the events are not due to bremsstrahlung, and instead speculate on
the kind of new physics they would imply.  There would
have to be a new spin zero neutral particle of mass $60$ GeV,
which we call $X^0$.
We find that there are only two reasonable scenarios:
the new particle ($X^0$), would have to couple to the $Z$ and to a
new $Z'$ with a coupling constant $\alpha_{XZZ'}$ which is $\gsim 1$,
where the $Z'\nu\nu$ coupling is suppressed;
or there would have to be a non-Higgs scalar $X'^0$ with a huge universal
$X'll$ coupling. The $X'll$
coupling could not be proportional to the lepton mass, or else
there would be far too many $\gg\tau\tau$ events (none have been seen yet) and
far too few $\gg ee$ events.
There would also have to be a
fairly large $ZXX'$ coupling: $\alpha_{ZXX'} \gsim 1/3$.
Our goal is not to promote either of these models (they are not even
formulated in a gauge invariant way!),
but rather to chart out what kinds of models
would need to be considered should the L3 data prove to be new physics.

\section {Possibilities}

We consider the various possibilities for new physics to explain the
four L3 events. There are two topologies through which the process
can proceed: the s channel
($ee \rightarrow A^{0(*)} \rightarrow X^0(\gg) B^{0*}(ll)$)
and t channel
($ee \rightarrow ee\, A^{0*}B^{0*} \rightarrow ee\, X^0(\gg)$)
(see figures 1 and 2).
By the well known result of Landau and Yang \cite{yang},
$X^0$ can only be a scalar or pseudoscalar.
The t channel is only important if $A$ and $B$ are photons.
As we will see, the s channel must have $A=Z$ and $B=Z'$ or $X'$,
where $Z'$ ($X'$) is a new spin 1 (spin 0) boson.

There are two problems associated with producing four \g\g\ll\ events:
producing enough $X^0$'s, and having $B(X\rightarrow\gg) \sim {\cal O}(1)$.
It is possible to construct a two Higgs doublet
model with the second property by
having one of the Higgs doublets decouple from fermions \cite{HHGaKANE},
though such a model would require a fine-tuning of order $10^{-3}$
\cite{barger}.
We will assume that either through some symmetry or fine tuning that
$B(X\rightarrow\gg) \sim {\cal O}(1)$, and concentrate on the first problem.

The production problem is much more difficult.
Consider the s channel.
Since LEP sits on the $Z$ pole, the largest cross section will come
for $A^0 = Z^0$.  For $B(X\rightarrow\gg) \sim {\cal O}(1)$, we
need $B(Z \rightarrow X B^*(ll))$ to be of order $2\times 10^{-6}$
(for each $l$).
A Standard Model (SM) $hZZ$ vertex gives a branching ratio only of order
$0.4 \times10^{-6}$ \cite{barger}, which is five times too small.
But we are not even allowed a diagonal coupling, because
if $B$ were also a $Z$, L3 would have seen about 12 $\gg \nu\bar\nu$
events \cite{barger}, which were not observed \cite{Lthree}.
Therefore \underbar{$B$ cannot be a $Z$}, and the $XAB$ coupling will
be off-diagonal.

There are then three possible ways to produce the $\gg ll$ events from
new physics:
via an $XZZ'$ or $ZXX'$ coupling through the s channel, or via
the t channel using only the $X\gg$ vertex, which would
be similar to producing a heavy pion.
The t channel production cross section of such a heavy pion is
enhanced by $|\ln(\sqrt{s}/2 m_e)|^2$ and
is proportional to $\Gamma(X\rightarrow \gg) / M_x^3$ \cite{Low}.
Such a heavy pion could produce several $\gg ee$ events (in 27 pb$^{-1}$)
if $\Gamma(X\rightarrow \gg)$ were a few hundred keV.
One might worry about three photon events  (with $M_{\gg} \simeq 60$GeV)
coming from the s channel,
which would have been found easily,
but it turns out that
a particle of this width would produce less than one event at LEP.
Unfortunately there is no way to generate enough
$\gg \mu\mu$ events.
If this were the mechanism, we would expect to
see many more $\gg ee$ than $\gg \mu\mu$ events,
contrary to observation, so the heavy pion scenario is ruled out.

The best possibility is to have $A=Z$ (on shell), and $B=Z'$.
This new $Z'$ boson
must couple weakly to neutrinos
(the ratio of $\nu$ to $l$ couplings squared must be about $1/12$ the
SM value),
and cannot mix too strongly with the
SM $Z$, or else we again get too many $\gg \nu\bar\nu$ events.
Further, the $XZZ$ diagonal coupling must be small
($\alpha_{XZZ} \sim \alpha_{1}$) for the same reason.
There are two new couplings which come into the $X^0$ production
cross section as depicted in figure 1: the $XZZ'$ coupling and the
$Z'll$ coupling, which we write as
\begin{eqnarray}
XZZ'&:\ &g_{XZZ'} M_Z' g^{\mu\nu}\nonumber \\
Z'll&:\ &{g_2 \over \cos\theta_W} \gamma^\mu(g_V(l)' + g_A(l)' \gamma^5)
\label{test}
\end{eqnarray}
where we have used the favorable value of $M_Z'$ for the off-diagonal
coupling.  We can use (\ref{test}) to write
\beq
B(Z\rightarrow X Z'^*(ll)) \sim g_{XZZ'}^2 \lambda_{Z'll}^2 ,
\eeq

\noi
where we define
\beq
\lambda_{Z'll}^2 \equiv
{ (g_2/\cos\theta_W)^2 (g_V(l)'^2 + g_A(l)'^2) \over    M_Z'^2} .
\eeq

We have neglected the momentum dependence of the propagator,
but that effect is small and tends to
cancel with that of the bounds on $\lambda_{Z'll}^2$.
In return, we have
put all dependence on $M_Z'$ and the couplings into one parameter which we
can constrain.
If we make the minimal assumptions about $Z'$ ($i.e.$ that the only
$Z'ff$ couplings which are necessarily non-zero are those of $Z'll$),
 we can get constraints
on $\lambda_{Z'll}^2$ only from leptonic processes
such as $e^+e^- \rightarrow l^+l^-$
and (g - 2) experiments.
The latter turns out to be a weaker constraint \cite{mery},
so we use data from TRISTAN
\cite{TRISexpt,TRIStheo}
on the former to obtain the conservative estimate
\begin{equation}
{\lambda_{Z'll}^2 \over \lambda_{Zll}^2} \lsim {1\over 4}.
\label{lambda ratios}
\end{equation}

\noi
Using this estimate we find that the off-diagonal $XZZ'$ coupling
constant must be greater than one (even though $\alpha_{XZZ} \sim \alpha_{1}$),
\begin{equation}
\alpha_{XZZ'} \equiv {g_{XZZ'}^2 \over 4 \pi} \gsim
 {1 \over B(X\rightarrow \gg)},
\end{equation}

\noi
implying that the theory is non-perturbative.  Such a large coupling could
also cause problems with the $\rho$ parameter, though this cannot be
addressed until one has a better understanding
of the underlying theory which includes $X$ and $Z'$.

The remaining possibility involves a second scalar $X'^0$ with unusual
couplings.
One cannot simply let $B=X$ because then $X$ couples strongly to fermions,
and $B(X \rightarrow \gg)$ is again small.  The $X'$ coupling
constant to charged leptons, $g_{X'll}$,
cannot be proportional to $m_l$, because then there would be too
many $\gg \tau\tau$ events and not enough $\gg ee$ events.
For a scalar lepton coupling, $g_{X'll}$
would need to be {\it huge}.  Because of this, $X'$
can actually give a fairly large contribution to (g - 2),
though the $e^+e^- \rightarrow l^+l^-$ constraint is still better.
Using TRISTAN data \cite{TRISexpt}, we find (using a rough estimate)
that $g_{X'll}^2 \lsim r'^2/50$, which  implies that

\begin{equation}
\alpha_{ZXX'} \gsim r'^2 /3,
\end{equation}

\noi
where $r' \equiv M_X' / M_Z.$  It is possible that this theory is
perturbative, depending upon what bounds one can put on $M_X'$ in the
full theory. We can at least say that $M_X' > 30$ GeV because
otherwise $X'$ could be on shell, and the leptons would have an
invariant mass equal to $M_X'$, which is not observed \cite{Lthree}.

Next we investigate the possible effect of $X'$ on the electron mass.
We will have to allow $X'^0$ to have a small vacuum expectation value (VEV)
at tree level so
as to allow for the counterterms we will need
to cancel the divergences coming from tadpole diagrams (see figure 3).
This could in principle give a contribution to $m_e$ larger than the
observed value.
For an estimate of the size of this effect, we take the renormalized VEV to be
just the finite size of the tadpole diagram in figure 3, which goes as
$g_{X'ff} m_f^3/M_X'^2$.
This implies that the
$X'tt$ coupling has to be negligible (or exactly zero) in order that
the contribution to the
electron mass, $\Delta m_e$, is smaller than $m_e$.
If only $g_{X'll}$ is non-zero, $\Delta m_e < 1$ keV.

\section{Concluding Remarks}

We have discussed possible models to explain the four L3 $\gg ll$ events
and concluded that the spectrum of choices is extremely limited, because
of the large production cross section for $X^0$, and because there have
been no $\gg \nu\bar\nu$ events seen.  There are still two classes of
models containing a $60$ GeV spin 0 neutral particle,
which are not immediately ruled out,
employing a new $Z'$ vector boson or new $X'$ non-Higgs scalar.

They both proceed through figure 1 with $A=Z,\ B=Z'$ and
$A=Z,\ B=X'$, respectively.
The heavy pion production scenario of figure 2 is ruled out
because it cannot produce enough $\gg\mu\mu$ events.
The $Z'$ must couple about twelve times more weakly to $\nu\nu$
relative to $ll$ than in the SM and cannot mix too strongly with the
SM $Z$.  This model has an $\alpha_{XZZ} \sim \alpha_1$, but  an
$\alpha_{XZZ'}$ which is $>1$, which implies the theory is non-perturbative.

The $X'$ must have non-Higgs-like $X'll$ couplings which are not proportional
to
fermion mass.  The theory may or may not be perturbative, depending
upon the size of $M_X'$.

Any complete theory based upon either model may fail to give a small
enough value for the $\rho$ parameter, or fail to maintain
$VV$ scattering unitarity.  If no $\gg\tau\tau$ events are found,
both theories would require non-universal generational couplings.
Discovery of a small number of $\gg\nu\nu$ events would not rule
out either model, but would point to a non-zero effective $XZZ$ or
$Z'\nu\nu$ coupling.  If more $\gg ff$ events
were found,
one could easily distinguish between the two classes of theories by looking
at the decay spectra of the $ff$ pair, since $Z'$ ($X'$) is
spin 1 (spin 0).

Although neither model is very attractive, and both still need to be formulated
in a gauge invariant way, we believe it is useful to know what kind of new
physics one has to consider to explain such a rich collection of
``rare $Z$'' events.




\end{document}